\documentstyle[12pt,epsf]{article}
\newcommand{\be}{\begin{equation}}
\newcommand{\ee}{\end{equation}}

\newcommand{\beqa}{\begin{eqnarray}}
\newcommand{\eeqa}{\end{eqnarray}}
\newcommand{\nn}{\nonumber}

\newcommand{\eqref}[1]{(\ref{#1})}

\def\boxit#1{\vbox{\hrule\hbox{\vrule\kern8pt
\vbox{\hbox{\kern8pt}\hbox{\vbox{#1}}\hbox{\kern8pt}}
\kern8pt\vrule}\hrule}}
\def\mathboxit#1{\vbox{\hrule\hbox{\vrule\kern8pt\vbox{\kern8pt
\hbox{$\displaystyle #1$}\kern8pt}\kern8pt\vrule}\hrule}}

\def\IB{\relax\hbox{$\inbar\kern-.3em{\rm B}$}}
\def\IC{\relax\hbox{$\inbar\kern-.3em{\rm C}$}}
\def\ID{\relax\hbox{$\inbar\kern-.3em{\rm D}$}}
\def\IE{\relax\hbox{$\inbar\kern-.3em{\rm E}$}}
\def\IF{\relax\hbox{$\inbar\kern-.3em{\rm F}$}}
\def\IG{\relax\hbox{$\inbar\kern-.3em{\rm G}$}}
\def\IGa{\relax\hbox{${\rm I}\kern-.18em\Gamma$}}
\def\IH{\relax{\rm I\kern-.18em H}}
\def\IK{\relax{\rm I\kern-.18em K}}
\def\IL{\relax{\rm I\kern-.18em L}}
\def\IP{\relax{\rm I\kern-.18em P}}
\def\IR{\relax{\rm I\kern-.18em R}}
\def\IZ{\relax\ifmmode\mathchoice
{\hbox{\cmss Z\kern-.4em Z}}{\hbox{\cmss Z\kern-.4em Z}}
{\lower.9pt\hbox{\cmsss Z\kern-.4em Z}} {\lower1.2pt\hbox{\cmsss
Z\kern-.4em Z}}\else{\cmss Z\kern-.4em Z}\fi}

\def\II{\relax{\rm I\kern-.18em I}}

\def\CN {{\cal N}}
\def\CO {{\cal O}}
\def\CP {{\cal P}}

\def\CS {{\cal S}}

\begin{document}

\hfill  CERN-PH-TH/2011-046

\hfill  NRCPS-HE/2011-010

\vspace{1cm}
\begin{center}
{\Large ~\\{\it  Extensions of the Poincar\'e Group

}

}

\vspace{2cm}

{\sl Ignatios Antoniadis$^{1}$\footnote{${}^\dagger$ On leave of absence
from CPHT {\'E}cole Polytechnique, F-91128, Palaiseau Cedex,France.} ,
Lars Brink$^{2}$ and  George Savvidy$^{13}$

\bigskip
\centerline{${}^1$ \sl Department of Physics, CERN Theory Division CH-1211 Geneva 23, Switzerland}
\bigskip
\centerline{${}^2$ \sl Department of Fundamental Physics, }
\centerline  {\sl  Chalmers University of Technology, S-412 96 G\"oteborg, Sweden}
\bigskip
\centerline{${}^3$ \sl Demokritos National Research Center, Ag. Paraskevi,  Athens, Greece}
\bigskip

}
\end{center}
\vspace{2cm}

\centerline{{\bf Abstract}}

\vspace{12pt}

\noindent
We construct an extension of the Poincar\'e group which involves
a mixture of internal and space-time supersymmetries. The resulting group is an
extension of the superPoincar\'e group with infinitely many generators which carry
internal and space-time indices.  It is a closed algebra since all Jacobi
identities are satisfied and it has therefore explicit matrix representations. We investigate the
massless case and construct the  irreducible representations of the extended symmetry. They
are divided into two sets, longitudinal and transversal representations.
The transversal representations involve an infinite series of  integer and half-integer helicities.
Finally we suggest an extension of the conformal group along the same line.


\newpage

\pagestyle{plain}


\section{\it Introduction}
In an attempt to discuss higher spin gauge fields in a new setting, a generalization of the Poincar\'e algebra has
been suggested \cite{Savvidy:2010vb}. In it the Poincar\'e generators are
enlarged by infinitely many new bosonic generators which carry
internal and space-time indices. In this article we shall construct a supersymmetric
extension of this algebra. The resulting algebra contains the ordinary superPoincar\'e generators together with
infinitely many bosonic generators which form a current algebra between themselves.
It is a closed algebra since all Jacobi identities are satisfied and it
can hence have explicit matrix representations.

Let us first introduce the infinite set of translationally invariant
generators which carry internal and space-time indices:
\be\label{represocurrentsubalgebra}
L_{a}^{\lambda_1 ... \lambda_{s}} ~,~~~~~~~s=0,1,2......
\ee
where $L_a~(s=0)$ are the generators of the internal Lie algebra $L_G$
and the generators $L_{a}^{\lambda_1 ... \lambda_{s}}$ are totally symmetric
with respect to the indices $  \lambda_1 ... \lambda_{s}  $.
These generators carry space-time and internal indices and  transform  under the
operations of both  groups. In a sense these generators remind us of gauge fields having
both Lorentz and internal indices and, as we shall see, there are some properties inherent of
gauge fields in them.
The current algebra of these generators
is defined as follows \cite{Savvidy:2010vb}:
\be\label{subalgebra1}
[L_{a}^{\lambda_1 ... \lambda_{n}}, L_{b}^{\lambda_{n+1} ... \lambda_{s}}]=if_{abc}
L_{c}^{\lambda_1 ... \lambda_{s} },~~~~s=0,1,2.....
\ee
where at the basic level (s=0) it contains the internal algebra $L_G$ with commutators $[L_a,L_b] =i f_{abc} L_c$.
The  current  algebra (\ref{subalgebra1}) is not yet completely defined  because
it does not specify how the new generators $L_a^{\lambda_1 ...  \lambda_{s}}$ transform under
space-time transformations. Assuming the generators $L_a^{\lambda_1 ...  \lambda_{s}}$ be translationally invariant
tensors of rank s, the following extension of
the Poincar\'e algebra  was  suggested \cite{Savvidy:2010vb}:
\beqa\label{gaugePoincare}
~&&[P^{\mu},~P^{\nu}]=0, \label{extensionofpoincarealgebra}\\
~&&[M^{\mu\nu},~P^{\lambda}] = i(\eta^{\lambda \nu}~P^{\mu}
- \eta^{\lambda \mu }~P^{\nu}) ,\nn\\
~&&[M^{\mu \nu}, ~ M^{\lambda \rho}] = i(\eta^{\mu \rho}~M^{\nu \lambda}
-\eta^{\mu \lambda}~M^{\nu \rho} +
\eta^{\nu \lambda}~M^{\mu \rho}  -
\eta^{\nu \rho}~M^{\mu \lambda} ),\nn\\
\nn\\
\label{levelcommutators}
~&&[P^{\mu},~L_{a}^{\lambda_1 ... \lambda_{s}}]=0,  \\
~&&[M^{\mu \nu}, ~ L_{a}^{\lambda_1 ... \lambda_{s}}] = i(
\eta^{\lambda_1\nu } L_{a}^{\mu \lambda_2... \lambda_{s}}
-\eta^{\lambda_1\mu} L_{a}^{\nu\lambda_2... \lambda_{s}}
+...+
\eta^{\lambda_s\nu } L_{a}^{\lambda_1... \lambda_{s-1}\mu } -
\eta^{\lambda_s\mu } L_{a}^{\lambda_1... \lambda_{s-1}\nu } ),\nonumber\\
\nn\\
\label{alakac}
~&&[L_{a}^{\lambda_1 ... \lambda_{n}}, L_{b}^{\lambda_{n+1} ... \lambda_{s}}]=if_{abc}
L_{c}^{\lambda_1 ... \lambda_{s} }       ~~~(  s=0,1,2,... ).
\eeqa
The first three commutators define the Poincar\'e
algebra as its subalgebra. The next two commutators tell us that the generators
$L_{a}^{\lambda_1 ... \lambda_{s}}$ are translationally invariant tensors of rank s
and the last commutator defines the current subalgebra (\ref{subalgebra1}).
One can check that all Jacoby identities are satisfied and we have an example
of fully consistent algebra, which is called an {\it extended Poincar\'e algebra $L_G(\CP)$
associated with a compact Lie group G.} Thus the algebra $L_G(\CP)$ incorporates
the Poincar\'e algebra and an internal algebra $L_G$ in a nontrivial way, which is different from
the direct product. The generators $L_{a}^{\lambda_1 ... \lambda_{s}}$
have a nonzero commutation relation with $  M^{\mu\nu} $ and therefore
carry higher spins.

\section{\it Supersymmetric Extension  of the $L_G(\CP)$ Algebra}

We are interested in constructing further extensions of the $L_G(\CP)$ algebra which should
include anti-commuting generators. {\it A priory} it is not obvious that such an extension
can be constructed. With this intention in mind, let us compare the above extension of the Poincar\'e algebra
with the (extended) super-Poincar\'e algebra which is defined as follows
\cite{Golfand:1971iw,Ramond:1971gb,Neveu:1971rx,Volkov:1973ix,Wess:1974tw,Haag:1974qh}:
\beqa\label{superNextensionofpoincarealgebra}
~&&[P^{\mu},~P^{\nu}]=0, \\
~&&[M^{\mu\nu},~P^{\lambda}] = i(\eta^{\lambda \nu}~P^{\mu}
- \eta^{\lambda \mu }~P^{\nu}) ,\nn\\
~&&[M^{\mu \nu}, ~ M^{\lambda \rho}] = i(\eta^{\mu \rho}~M^{\nu \lambda}
-\eta^{\mu \lambda}~M^{\nu \rho} +
\eta^{\nu \lambda}~M^{\mu \rho}  -
\eta^{\nu \rho}~M^{\mu \lambda} ), \nn\\
\nn\\
\label{superNextensionofpoincarealgebra1}
~&&[P^{\mu},~Q^{i}_{\alpha}]=0, \\
~&&[M^{\mu \nu}, ~ Q^{i}_{\alpha}] = {i\over 2} (
\gamma^{\mu\nu} Q^{i})_{\alpha} ,~~~~~~
\gamma^{\mu\nu}={1\over 2} [\gamma^{\mu},\gamma^{\nu}]  \nn\\
\nn\\
\label{5}
~&&\{ Q^{i}_{\alpha}, Q^{j}_{\beta} \}=-2~\delta^{ij} (\gamma^{\mu} C)_{\alpha\beta}
 P_{\mu},~~~~~i=1,...,N,
\eeqa
where we allowed for an $R$-symmetry specified by the indices $i$ and $j$. $Q^{i}_{\alpha}$ is a Majorana spinor.
This algebra also has  the Poincar\'e algebra (\ref{gaugePoincare}), (\ref{superNextensionofpoincarealgebra})
as a subalgebra.  The next two commutators (\ref{levelcommutators})
and (\ref{superNextensionofpoincarealgebra1}) express the fact
that the extended generators $Q^{i}_{\alpha}$ and $L_{a}^{\lambda_1 ... \lambda_{s}}$
are translationally invariant operators and carry a nonzero spin.
The last commutators (\ref{alakac}) and (\ref{5}) are
essentially different in both algebras: in super-Poincar\'e algebra the generators
$Q^{i}_{\alpha}$ anti-commute with the operator $P^{\mu}$, while in our
case $L_{a}^{\lambda_1 ... \lambda_{s}}$ commute with themselves to form an infinite series
of commutators of the current algebra (\ref{subalgebra1}) which cannot be truncated.
Therefore, the index $s$ runs from zero to infinity, providing an example
of an infinitely-dimensional current subalgebra   \cite{Faddeev:1984jp}.

Another possibility to combine the above algebras would be to consider an
infinite set of spinor-tensor generators $Q^{i}_{\alpha \lambda_1 ... \lambda_{s}}$,
but this does not work. Therefore the natural suggestion is the following unification of these
algebras\footnote{Our notational conventions follow the article \cite{Derendinger:1990tj}.}:
\beqa\label{supergaugePoincare}
~&&[P^{\mu},~P^{\nu}]=0, \label{extensionofpoincarealgebra}\\
~&&[M^{\mu\nu},~P^{\lambda}] = i(\eta^{\lambda \nu}~P^{\mu}
- \eta^{\lambda \mu }~P^{\nu}) ,\nn\\
~&&[M^{\mu \nu}, ~ M^{\lambda \rho}] = i(\eta^{\mu \rho}~M^{\nu \lambda}
-\eta^{\mu \lambda}~M^{\nu \rho} +
\eta^{\nu \lambda}~M^{\mu \rho}  -
\eta^{\nu \rho}~M^{\mu \lambda} ),\nn\\
\nn\\
\label{levelcommutatorsss}
~&&[P^{\mu},~L_{a}^{\lambda_1 ... \lambda_{s}}]=0,  \\
~&&[P^{\mu},~Q^{i}_{\alpha}]=0, \nn\\
~&&[M^{\mu \nu}, ~ L_{a}^{\lambda_1 ... \lambda_{s}}] = i(
\eta^{\lambda_1\nu } L_{a}^{\mu \lambda_2... \lambda_{s}}
-\eta^{\lambda_1\mu} L_{a}^{\nu\lambda_2... \lambda_{s}}
+...+
\eta^{\lambda_s\nu } L_{a}^{\lambda_1... \lambda_{s-1}\mu } -
\eta^{\lambda_s\mu } L_{a}^{\lambda_1... \lambda_{s-1}\nu } ),\nonumber\\
~&&[M^{\mu \nu}, ~ Q^{i}_{\alpha}] = {i\over 2} (
\gamma^{\mu\nu} Q^{i})_{\alpha} ,~~~~~~
\gamma^{\mu\nu}={1\over 2} [\gamma^{\mu},\gamma^{\nu}]  \nn\\
\nn\\
\label{alakacss}
~&&[L_{a}^{\lambda_1 ... \lambda_{n}}, L_{b}^{\lambda_{n+1} ... \lambda_{s}}]=if_{abc}
L_{c}^{\lambda_1 ... \lambda_{s} }     ,  ~~~  s=0,1,2,...   \nn\\
~&&\{ Q^{i}_{\alpha},~ Q^{j}_{\beta} \}=-2~\delta^{ij} (\gamma^{\mu} C)_{\alpha\beta}
 P_{\mu},~~~~~i=1,...,N  \nn\\
~&&[L_{a}^{\lambda_1 ... \lambda_{s}}, Q^{i}_{\alpha} ]=0~.
\eeqa
Here, at $s=0$, we have the relations
$$
[P^{\mu},~L_{a} ]=0,~~~[M^{\mu \nu}, ~ L_{a} ] =0 ,
$$
therefore the internal bosonic  algebra $L_G$ obeys the Coleman-Mandula theorem \cite{Coleman:1967ad}.

Let us now investigate the commutators between $Q^{i}_{\alpha}$ and the rest of the generators.
First we have to check the Jacobi identities which contain at least one anticommuting
generator. They are:
\beqa
~&&[~[L_{a}^{\lambda_1 ... \lambda_{s}},~~~ P^{\mu}] ~~~~~Q^{i}_{\alpha}]+ Perm. =0\nn\\
~&&[[L_{a}^{\lambda_1 ... \lambda_{s}},~~~ M^{\mu\nu}]~~~~ Q^{i}_{\alpha}]+ Perm.=0\nn\\
~&&[[L_{a}^{\lambda_1 ... \lambda_{n}},~ L_{b}^{\lambda_{n+1} ... \lambda_{s}}]~Q^{i}_{\alpha}]+ Perm.=0\nn
\eeqa
and as one can check, they are indeed identically true. The identities with two anticommuting generators have the form
\beqa
\{ ~[L_{a}^{\lambda_1 ... \lambda_{s}},~Q^{i}_{\alpha}]~Q^{j}_{\beta} \}+
\{ ~[L_{a}^{\lambda_1 ... \lambda_{s}},~Q^{j}_{\beta}  ]~Q^{i}_{\alpha} \}+
[\{ ~Q^{i}_{\alpha},~Q^{j}_{\beta} \}]~~~L_{a}^{\lambda_1 ... \lambda_{s}}]=0\nn
\eeqa
and  they are also true. The rest of the identities are satisfied since they coincide with the identities of the known
subalgebras (\ref{gaugePoincare})-(\ref{alakac}) and (\ref{superNextensionofpoincarealgebra})-(\ref{5}).

\section{\it General Properties of the Extended  Algebra $L_G(\CS\CP)$}

The algebra (\ref{supergaugePoincare}) is  invariant with respect to the following "gauge" transformations:
\beqa\label{isomorfism}
&& L_{a}^{\lambda_1 ... \lambda_{s}} \rightarrow L_{a}^{\lambda_1 ... \lambda_{s}}
+ \sum_{1} P^{\lambda_1}L_{a}^{\lambda_2 ... \lambda_{s}}+
\sum_{2} P^{\lambda_1} P^{\lambda_2} L_{a}^{\lambda_3 ... \lambda_{s}} +...+
P^{\lambda_1}... P^{\lambda_s} L_{a}\nn\\
&&P^{\lambda} ~~~\rightarrow ~~~P^{\lambda},\\
&& M^{\mu\nu}~~ \rightarrow ~~ M^{\mu\nu},\nn\\
&& Q^{i}_{\alpha}~~ \rightarrow ~~Q^{i}_{\alpha},\nn
\eeqa
where the sums  $\sum_{1},\sum_{2},... $ extend over all inequivalent index permutations.
It is not an internal isomorphism since it cannot be represented as
conjugations by elements $U$ of the group  itself: $L \rightarrow U^{-1} ~L ~U$.
The transformations contain polynomials of the commuting momenta and are reminiscent
of the gauge transformations for the gauge fields. They are ``off-shell" transformations
because the invariant operator $P^2$ can have any value\footnote{
Note that the square mass operator $P^2$,~is a Casimir invariant for the above algebra
while the spin operator $W^{\mu}W_{\mu}$ ($W^{\mu}$ being the Pauli-Lubansky vector) is not.}.
As a result, to any given  representation of
$L_{a}^{  \lambda_1 ... \lambda_{s}},~s=1,2,...$  of the extended algebra
one can add the longitudinal terms, as it follows from the  transformation (\ref{isomorfism}).
Thus  all representations are defined modulo "gauge transformations"
and we can identify these generators as "gauge generators".
\\
\\
Theorem. {\it To any given  representation of the gauge generators  $L_{a}^{  \lambda_1 ... \lambda_{s}},~s=1,2,...$
of the extended algebra  one can add longitudinal terms.
  All representations
are therefore defined modulo longitudinal terms.}
\\

The second general property of the extended algebra is that each gauge generator
$ L_{a}^{  \lambda_1 ... \lambda_{s}}$ cannot be
realized as an irreducible representation of the Poincar\'e subalgebra of a definite helicity,
i.e. to be a symmetric and {\it traceless tensor}. The reason for this is that the commutator of
two symmetric traceless generators in the current subalgebra (\ref{subalgebra1}) is not any more
a traceless tensor. Therefore the gauge generators should realize a reducible representation
of the Poincar\'e subalgebra and each of them carries a  sequence of helicities, which we shall
find out in the subsequent sections.

Finally, the extended algebra $L_G(\CS\CP)$ has a general reducible
representation in terms of differential operators of the following
form:
\beqa\label{represofextenpoincarealgebra}
~&& P^{\mu} = k^{\mu} ,\nn\\
~&& M^{\mu\nu} = i(k^{\mu}~ {\partial\over \partial k_{\nu}}
- k^{\nu }~ {\partial \over \partial k_{\mu}}) + i(\xi^{\mu}~ {\partial\over \partial \xi_{\nu}}
- \xi^{\nu }~ {\partial \over \partial \xi_{\mu}}) - {i \over 2} \bar{\vartheta}
 \gamma^{\mu\nu}  {\partial \over \partial \vartheta },\nn\\
~&& Q_{\alpha}  = -i{\partial \over \partial \bar{\vartheta}_{\alpha}} +i
  ( \gamma^{\mu} \vartheta)_{\alpha }  k_{\mu},\\
~&& L_{a}^{\lambda_1 ... \lambda_{s}} =\xi^{\lambda_1}...\xi^{\lambda_s} \otimes L_a ,\nn
\eeqa
where the vector superspace of functions is parameterized
in terms of momentum coordinates $k^{\mu}$, translationally invariant vector variables $\xi^{\mu}$
and  anticommuting Grassmann variables $\vartheta_{\alpha}$
\be\label{vectorspace}
\Psi(k^{\mu}, \xi^{\nu}, \vartheta_{\alpha})~.
\ee
This representation allows us to further justify
the interpretation of the  transformation (\ref{isomorfism}) as a gauge transformation and of the generators
 $ L_{a}^{  \lambda_1 ... \lambda_{s}}$  as gauge generators if one considers how this transformation acts
on the representation (\ref{represofextenpoincarealgebra}). Indeed, the transformation (\ref{isomorfism})
 induces a transformation for
the vector variable  $\xi^{\mu}$ of the form
\be\label{gaugetransform}
\xi^{\mu} \rightarrow \xi^{\mu} +    k^{\mu},
\ee
reminiscent of a gauge transformation for the photon polarization vector. Furthermore
in order to obtain the irreducible representations from (\ref{represofextenpoincarealgebra}),
we shall follow  Wigner's prescription imposing  invariant constraints on the
vector space of functions defined in (\ref{vectorspace})
of the following form \cite{wigner1,yukawa1}:
\be\label{constraint}
k^2=0,~~~k^{\mu} \xi_{\mu}=0,~~~\xi^2=-1~.
\ee
These equations have a unique solution
\be\label{solution}
\xi^{\mu}= \xi k^{\mu} + e^{\mu}_{1}\cos\varphi +e^{\mu}_{2}\sin\varphi,
\ee
where $e^{\mu}_{1}=(0,1,0,0),~ e^{\mu}_{2}=(0,0,1,0)$ when $k^{\mu}=k(1,0,0,1)$,
thus justifying the interpretation of the vector variable $\xi^{\mu}$ as a polarization
vector\footnote{In this article we shall consider only massless representations with $k^2=0$ .}. The
invariant subspace of functions is now reduced to the form $\Psi(k^{\mu}, \xi , \varphi, \vartheta_{\alpha})$,
where $\xi$ and $\varphi$ remain as independent variables.

There are important properties of the above representation (\ref{represofextenpoincarealgebra}),
(\ref{constraint}) and (\ref{solution}) which are worth mentioning:

$(i)$ The gauge transformation
(\ref{isomorfism}), (\ref{gaugetransform}) cannot trivialize the above representation by nullifying the
generators $L_{a}^{\lambda_1 ... \lambda_{s}}$, but what it can do is to change the parameter $\xi$ in front
of $k^{\mu}$ in (\ref{solution})  and

$(ii)$  This representation is transversal in the sense that
\be
k_{\lambda_1}L_{a}^{\lambda_1 ... \lambda_{s}}=0,~~~~s=1,2,...
\ee
Having in hand this interpretation of the generators $L_{a}^{\lambda_1 ... \lambda_{s}}$ we can
divide the vector space of representations into {\it pure longitudinal} and {\it transversal} subsets.

\section{\it Longitudinal Representations}
Let us consider an irreducible representation of the
superPoincar\'e algebra (\ref{superNextensionofpoincarealgebra}), in which the generators
$P^{\mu}, M^{\mu\nu}, Q^{i}_{\alpha}$ realize a matrix representation with maximal helicity $h$
and  the $L_a$  realize an irreducible matrix representation of the internal algebra $L_G$.
If one now takes the gauge generators in the trivial form $L_{a}^{  \lambda_1 ... \lambda_{s}} = 0,~s=1,2,...$ it is easy to check
that this set of generators fulfils all commutation relations of the
algebra (\ref{supergaugePoincare}) and therefore forms a true representation
of the extended algebra $L_G(\CS\CP)$. Applying the above theorem to the representation
just described
we find that it is isomorphic to the representation in which all generators remain in the same
matrix form, except that the gauge generators $L_{a}^{  \lambda_1 ... \lambda_{s}}$
are now purely longitudinal. Thus we have the following equivalence relation:
\beqa
L_{\CS\CP}:~&P^{\mu}, ~ M^{\mu\nu},~Q^{i}_{\alpha} &~~~~~~~~~~~~~~~~~~~~~~~~~~~~~~~~~~~~P^{\mu}, ~M^{\mu\nu} ,~Q^{i}_{\alpha} \nn \\
&~~~~L_{a}^{  \lambda_1 ... \lambda_{s}} = 0,&~~~~~~~~~~\Leftrightarrow~~~~~~~~~~~~~~~~~~~~~
L_{a}^{ \vert \vert  \lambda_1 ... \lambda_{s}}=
k^{\lambda_1}... k^{\lambda_s} \oplus L_{a}\nn\\
L_G:~&~L_a  &~~~~~~~~~~~~~~~~~~~~~~~~~~~~~~~~~~~~~~~~   L_a
\eeqa
where $s=1,2,..$ It states that representations with trivial generators $L_{a}^{  \lambda_1 ... \lambda_{s}} =0$
and representations with purely longitudinal generators
$L_{a}^{ \vert \vert  \lambda_1 ... \lambda_{s}} =k^{\lambda_1}... k^{\lambda_s} \oplus L_{a} $
are isomorphic to each other. In other words, pure longitudinal
representations   factorize  into super-Poincar\'e  $L_{\CS\CP}$ and internal $L_G$ algebra  multiplets.
Or, if one reads this statement from right to left,
it says that pure longitudinal representations of $L_{a}^{ \lambda_1 ... \lambda_{s}}$
carry no more helicities than the ones carried by the representation of the SuperPoincar\'e subgroup,
since it is equivalent to a trivial representation of $L_{a}^{ \lambda_1 ... \lambda_{s}}$, namely
$L_{a}^{ \lambda_1 ... \lambda_{s}} =0$ ($s=1,2,..$).

\section{\it Transversal Representations}

As we have seen in the previous section any representation of the extended algebra (\ref{supergaugePoincare})
in which the generators of the superPoincar\'e subalgebra (\ref{superNextensionofpoincarealgebra})
realize a matrix representation of finite multiplicity  is always equivalent to a representation in which the
gauge generators are longitudinal and therefore trivial. It seems natural to
think that in order to get a nontrivial representation
for the gauge generators one should consider infinite-dimensional representations of the superPoincar\'e
subalgebra (\ref{superNextensionofpoincarealgebra}). Such representations have been constructed in the
article \cite{Brink:2002zx}.

The irreducible representation of the extended algebra can be found by the well-known  method
of induced representations \cite{wigner,Brink:2002zx}. This method consists of finding a representation
of the Wigner's  little group L  and boosting it up to a representation
of the full group. The subgroup L is a group of transformations which leave
a fixed momentum, in our case time-like momentum $k^{\mu}=k(1,0,0,1)$, invariant.
The Poincar\'e generators in L form the Euclidean  algebra $E(2)$ (see Appendix for definitions)
\beqa
[h,\pi^{'}]=
i\pi^{''},~~~[h,\pi^{''}]= -i\pi^{'},~~~[\pi^{'},\pi^{''}]=0.\nn
\eeqa
Notice that transformations generated by the gauge $L_{a}^{ \lambda_1 ... \lambda_{s}}$
and supercharge $Q_{\alpha}$ generators
leave the manifold of states with fixed
momentum  invariant, since they all commute with $P^{\mu}$, therefore all these generators
should be included into the little algebra L, so that we have the following
generators in L\footnote{In this section we shall use two component Weyl spinors and only discuss $N=1$ supersymmetry.}:
\be\label{theLset}
h,~~~\pi^{'} ,~~~\pi^{''},~~~~Q_{\alpha}~,~~~~\bar{Q}_{\dot{\alpha}} ~,~~L_{a}^{ \lambda_1 ... \lambda_{s}}.
\ee
The full set of commutators of the L algebra are presented in the Appendix and have the following
form\footnote{Not all of them are presented here in the main text.}:
\beqa\label{superLittlealgebrass}
&&[h,\pi^{'}]=+
i\pi^{''},~~~~~~~~~[h,\pi^{''}]= -i\pi^{'},~~~~~~~~~~~~~~~[\pi^{'},\pi^{''}]=0,\nn\\
&&[h, ~ \bar{Q}_{\dot{1}} ] = -{1\over 2} \bar{Q}_{\dot{1}} ,~~~~~~[h, ~ Q_1 ]
=  +{1\over 2} Q_1,~~~~~~~~~~~\{Q_1, \bar{Q}_{\dot{1}}\}=4k\\
&&[\pi^{'}, Q_1] =iQ_2,~~ [\pi^{''}, Q_1] = -Q_2,~~[\pi^{'}, \bar{Q}_{\dot{1}}] =i\bar{Q}_{\dot{2}},~~
 [\pi^{''}, \bar{Q}_{\dot{1}}] = \bar{Q}_{\dot{2}}.~~~~\nn
\eeqa
The supercharges commute with the gauge generators
\be\label{gaugesupercharges}
~[Q_1,L^{\lambda}_{a}] = 0,~~~~~[\bar{Q}_{\dot{1}},L^{\lambda}_{a}] =0.
\ee
The commutators between the E(2) and the $L^{\lambda_1}_{a}$ generators are:
\beqa\label{extendedalgebra0}
\begin{array}{lll}
~[h, ~ L^{0}_{a} ] = [h, ~ L^{3}_{a} ]=0 ~~~~&~[\pi^{'}, ~ L^{0}_{a} ] =-i L_{a}^{1 } ~~~&[\pi^{''}, ~ L^{0}_{a} ] =-i L_{a}^{2 } \\
~[h, ~ L^{0}_{a} ] = [h, ~ L^{3}_{a} ]=0  &~[\pi^{'} , ~ L^{3}_{a} ]=-i L_{a}^{1 } ~~~&[\pi^{''} , ~ L^{3}_{a} ]=-i L_{a}^{2 }    \\
~[h, ~ L_{a}^{1 }] = +i L_{a}^{2 } &~[\pi^{'}, ~ L_{a}^{1 }] = -i (L^{0}_{a}-L_{a}^{3 }) ~~~&[\pi^{''}, ~ L_{a}^{1 }] = 0 \\
~[h, ~ L_{a}^{2 }] = -i L_{a}^{1 } &~ [\pi^{'}, ~ L_{a}^{2 }] = 0 ~~~&[\pi^{''}, ~ L_{a}^{2 }] = -i (L^{0}_{a}-L_{a}^{3 }) \\
\end{array}
\eeqa
and the higher rank generators $L_{a}^{ \lambda_1 ... \lambda_{s}}$ have similar structure of commutators (see
details in the Appendix).
The problem reduces to the construction of the unitary irreducible representations of the L  algebra.

The representation of the little  algebra L can be found by restricting the general
representation (\ref{represofextenpoincarealgebra}) into the invariant subspace defined by the
conditions (\ref{constraint}) and solution (\ref{solution}) to be of the form:
\be\label{independentvariables}
\Psi(k^{\mu}, \xi^{\nu}, \vartheta_{\alpha})~\delta(k^2)~\delta(k\cdot\xi)~\delta(e^2 +1)
= \Phi(k^{\mu}, \varphi, \xi, \vartheta_{\alpha}).
\ee
Making use of the chain rule we may reexpress (\ref{represofextenpoincarealgebra}) as a differential
operator in the new variables, so that the generators of the $L$ algebra reduce to the form
\beqa
&h= -i{\partial \over \partial\varphi} -
{1\over 2}(\vartheta^{1} { \partial \over \partial \vartheta^{1} }-
\bar{\vartheta}^{\dot{1}} {\partial \over \partial\bar{\vartheta}^{\dot{1}} }-
\vartheta^{2} { \partial \over \partial \vartheta^{2} }+
\bar{\vartheta}^{\dot{2}} {\partial \over \partial\bar{\vartheta}^{\dot{2}} }), \nn\\
&\pi^{'}= \rho \cos\varphi -i\vartheta^{1} { \partial \over \partial \vartheta^{2} }
-i\bar{\vartheta}^{\dot{1}} {\partial \over \partial\bar{\vartheta}^{\dot{2}}},~~~~
\pi^{''}=\rho \sin\varphi + \vartheta^{1} { \partial \over \partial \vartheta^{2} }
-\bar{\vartheta}^{\dot{1}} {\partial \over \partial\bar{\vartheta}^{\dot{2}}},~~~~
\rho = -{i\over k} {\partial \over \partial \xi},
\nn\\
&Q_{1}  =-i{ \partial \over \partial \vartheta^{1} } -2i k \bar{\vartheta}^{\dot{1}},~~~~~
Q_{2}  =-i { \partial \over \partial  \vartheta^2 },  \nn\\
&\bar{Q}_{\dot{1}}  = +i{ \partial \over \partial \bar{\vartheta}^{\dot{1}} } +2i k  \vartheta^{1},~~~~~
\bar{Q}_{\dot{2}}  = +i { \partial \over \partial  \bar{\vartheta}^{\dot{2}} }
\eeqa
and taking into account (\ref{solution}) the gauge generators become
\be\label{trasversalgenera}
L_{a}^{\bot~ \mu_1 ... \mu_{s}}= \prod^{s}_{i=1} ( \xi k^{\mu_i} + e^{\mu_i}_{1}\cos\varphi
+e^{\mu_i}_{2}\sin\varphi)\oplus L_a.
\ee
This is a purely transversal representation and, as we have already mentioned in the previous sections,
it cannot be trivialized by the transformations (\ref{isomorfism}), (\ref{gaugetransform}). It
is transversal in the sense that
\be
k_{\lambda_1}L_{a}^{\bot \lambda_1 ... \lambda_{s}}=0.~~~~s=1,2,...
\ee
What is important to notice is that the commutators between the generators of $E(2)$ and the $L_{a}^{\bot~ \lambda_1 ... \lambda_{s}}$
generators of the little algebra L are fulfilled only if $\rho \neq 0$. An example of this
is the commutator $[\pi^{'}, ~ L^{0}_{a} ] =-i L_{a}^{1 }$.

Next we are interested in knowing
the helicity content of the transversal gauge generators just constructed.
The supercharges $Q_1,  \bar{Q}_{\dot{1}} $ carry helicities $h=(1/2,-1/2)$, as one can see  from the commutators of the helicity operator with supercharges
in (\ref{superLittlealgebrass}). The Poincar\'e generators $\pi^{\pm} = \pi^{'} \pm \pi^{''}$ carry
helicities $h=(1,-1)$. The fact that the $L^{\pm}_a= L^{1}_{a} \pm i L^{2}_{a}$ carry helicities
$h=(1,-1)$ is seen from the commutators in the first column of (\ref{extendedalgebra0}):
\be
[h,~L^{\pm}_a] = \pm L^{\pm}_a.
\ee
The rank-2 generators $L^{++}_a, L^{+-}_a,L^{--}_a$ carry helicities $h=(2,0,-2)$, where
$$
L^{++}_a = L^{11}_a +2i L^{12}_a - L^{22}_a,~~~~~L^{+-}_a = L^{11}_a + L^{22}_a,~~~~~
L^{--}_a = L^{11}_a -2i L^{12}_a -L^{22}_a,
$$
so that $[h, ~L^{\pm\pm}_a] = \pm 2  L^{\pm\pm}_a,~~[h, ~L^{+-}_a]=0 $ and in general the rank-s
$(L^{+\cdot\cdot\cdot+}_{a},...,L^{-\cdot\cdot\cdot-}_{a})$ generators  carry helicities in the following range:
\be\label{trasversalgenera1}
h=(s,s-2,......, -s+2, -s),
\ee
in total $s+1$ states. (Remember that gauge generator
$ L_{a}^{  \lambda_1 ... \lambda_{s}}$ cannot be
realized as an irreducible representation of the Poincar\'e subalgebra of a definite helicity.)
This can be seen  also from the explicit representation (\ref{trasversalgenera}):
\be\label{trasversalgenera1}
L_{a}^{\bot~ \mu_1 ... \mu_{s}}= \prod^{s}_{n=1} ( \xi k^{\mu_n} + e^{i \varphi} e^{\mu_n}_{+}
+e^{-i \varphi} e^{\mu_n}_{-})\oplus L_a,
\ee
where $e^{\mu}_{\pm}= (e^{\mu}_1 \mp i e^{\mu}_2)/2$. The last formula also illustrates the
realization of the transformation rule (\ref{isomorfism}). Indeed if we perform the multiplication
in (\ref{trasversalgenera1}) and collect terms with a given power of momentum we get the
following expression
\beqa\label{trasversalgenera2}
L_{a}^{\bot~ \mu_1 ... \mu_{s}}= \prod^{s}_{n=1} (e^{i \varphi} e^{\mu_n}_{+}
+e^{-i \varphi} e^{\mu_n}_{-})\oplus L_a +~~~~~~~~~~~~~~~~~~~~~~~~~~~~~~~~~~~~~~~~~~~~~~ \\
+\sum_{1} \xi k^{\lambda_1} \prod^{s-1}_{n=1} (e^{i \varphi} e^{\mu_n}_{+}
+e^{-i \varphi} e^{\mu_n}_{-})\oplus L_a +...+\xi
k^{\lambda_1}... \xi k^{\lambda_s}\oplus  L_{a} ,\nn
\eeqa
where
\be
\prod^{s}_{n=1} (e^{i \varphi} e^{\mu_n}_{+}
+e^{-i \varphi} e^{\mu_n}_{-})\oplus L_a
\ee
is the transversal part of the generator which we describe in terms
of $(L^{+\cdot\cdot\cdot+}_{a},...,L^{-\cdot\cdot\cdot-}_{a})$.
The rest of the terms (corresponding to the terms with indices $0$ or $3$) are purely
longitudinal, transforming under (\ref{isomorfism}), and can be gauged away\footnote{The situation is analogous
to the polarization tensor of the graviton $e^{\mu\nu}(k)= e^{\mu\nu}_{1}  + e^{\mu\nu}_{2} + k^{\mu}\xi^{\nu}
+ k^{\nu}\xi^{\mu}$. The first two terms describe transversal polarizations, the last two terms describe
the longitudinal part and if one takes $\xi^{\mu} =e^{\mu}_{1}$ then there will be a spin one
part, but still representing a pure gauge.}.

The states of the representation can be constructed by using these operators. The massless irreducible
representation of $\CN=1$ supersymmetry comprises the two states with helicities $\lambda$ and
$\lambda -1/2$:
\beqa
\begin{array}{ll}
|\lambda>       &~~~~~~~\bar{Q}_{\dot{1}} |\lambda>  \\
\lambda &~~~~~~\lambda-1/2          ,
\end{array}
\eeqa
where $h |\lambda> = \lambda |\lambda>$ and  $Q_1 |\lambda> =0$.  Because the operators $\pi^{\pm}$ commute
with the supercharges
(\ref{superLittlealgebrass})\footnote{This is because on the state $ \vert \lambda >$ the supercharges $Q_2$,$\bar{Q}_{\dot{2}}$
are realized  trivially $Q_2 \vert \lambda > = \bar{Q}_{\dot{2}} \vert \lambda >=0$.}
they generate an infinite tower of high helicity states:
\beqa\label{supermultiples}
\begin{array}{lllll}
~~~~~~~....&~~\pi^{+} |\lambda>     &~~~~~~|\lambda>             &~~~~~~~~~\pi^{-} |\lambda>    &.... \\
~~~~~~~....&\pi^{+} \bar{Q}_{\dot{1}}|\lambda>     &~~~\bar{Q}_{\dot{1}}|\lambda>             &~~~~~~~\pi^{-}\bar{Q}_{\dot{1}} |\lambda>    &.... \\
\\
~~~~....&\lambda+1             &~~~~~~~~\lambda               &~~~~~~~\lambda -1           &..... \\
~~~~~~....&\lambda+1/2           &~~~~~~\lambda -1/2          &~~~~~~~\lambda -3/2  &.....
\end{array}
\eeqa
From the above formulae it follows that the infinite multiplets built up by any integer $\lambda$ are isomorphic to each other.
The same is true for multiplets built up by any half-integer $\lambda$. The supersymmetry transforms simultaneously
different pairs of states within the large multiplet, the vertical columns in (\ref{supermultiples}). It does not
transform nontrivially the whole multiplet, that is the horizontal states in (\ref{supermultiples}).
The operators $(L^{+\cdot\cdot\cdot+}_{a},...,L^{-\cdot\cdot\cdot-}_{a})$ commute with the supercharges
(\ref{gaugesupercharges}) and with the $\pi^{\pm}$ generators (\ref{extendedalgebra0})
and are similar to the creation and annihilation operators
of the Kac-Moody algebra. Therefore the state  $|\lambda>$
must also form an irreducible representation of the internal algebra $L_G$ from which color states of high helicity
are generated.

We note that these infinite representations are the same as those found in \cite{Brink:2002zx}, where the continuous spin
representations of the superPoincar\'e group were derived.

\section{\it Generalization of de Sitter and Conformal Groups}
We might ask if the extension above can be made for the de Sitter and the conformal groups, too.
Consider first the algebras $SO(4,1)$ or $SO(3,2)$
\beqa
[J^{AB}, J^{CD}]= i (g^{AD}J^{BC}-g^{AC}J^{BD} + g^{BC}J^{AD} -g^{BD}J^{AC}),\nn
\eeqa
where $g^{AB}=(+----)$ or $g^{AB}=(+---+)$  and A,B=0,1,..,4. The Wigner-In\"on\"u contraction
$J^{4\mu}=RP^{\mu}$, $J^{\mu\nu}=M^{\mu\nu}$,
where $\mu,\nu =0,1,2,3$ and  $R \rightarrow \infty$, reduces it to the Poincar\'e algebra.
In the previous analysis there were no restrictions on the dimension of space-time
when we considered the bosonic part. We can then drop the translation generators and just consider
 the sets of commutators
\beqa
~&&[J^{AB}, J^{CD}]= i (g^{AD}J^{BC}-g^{AC}J^{BD} + g^{BC}J^{AD} -g^{BD}J^{AC}),\nn\\
~&&[J^{AB}, ~ L_{a}^{C_1 ... C_{s}}] = i(
\eta^{C_1B } L_{a}^{A C_2... C_{s}}
-...-
\eta^{C_s A} L_{a}^{C_1... C_{s-1}B } ),\nonumber\\
~&&[L_{a}^{C_1 ... C_{n}}, L_{b}^{C_{n+1} ... C_{s} }]=if_{abc}
L_{c}^{C_1 ... C_{s} }       ~~~(  s=0,1,2,... ).
\eeqa
This is an obvious generalization to the cases of the (anti)de Sitter groups.

Similarly since  the $SO(d,2)$ algebra is isomorphic to the conformal algebra the
algebra $L_{G}(\CP)$ can be extended to the conformal group as well with the following well known identification:
\be
J^{\mu\nu}= M^{\mu\nu},~~~J^{\mu,d}= {1 \over 2}(K^\mu - P^\mu),~~~J^{\mu(d+1)}={1 \over 2}(K^\mu + P^\mu),~~~
J^{(d+1)d}=D,
\ee
where $g^{AB}=(+---...-+)$  and $A,B=(0,...,d,d+1)$. Thus we have the algebra $L_G(\CS\CO)$ of the
form
\beqa
~&&{1\over i}[J^{AB}, J^{CD}]=  g^{AD}J^{BC}-g^{AC}J^{BD} + g^{BC}J^{AD} -g^{BD}J^{AC}, \nn\\
~&&{1\over i}[J^{AB}, ~ L_{a}^{D_1 ... D_{s}}] =
\eta^{D_1B } L_{a}^{A D_2... D_{s}}
-...-
\eta^{D_s A} L_{a}^{D_1... D_{s-1}B } ,\nonumber\\
~&&{1\over i}[L_{a}^{D_1 ... D_{n}}, L_{b}^{D_{n+1} ... D_{s} }]= f_{abc}
L_{c}^{D_1 ... D_{s} }       ~~~~~~~~~~(  s=0,1,2,... ).
\eeqa
We defer to future work the study of generalisations of these algebras to superalgebras.

\section{\it Conclusions}

In this paper we have studied infinite-component massless supermultiplets which arise from a new extension of
the  superPoincar\'e algebra. We find that they agree with the continuous spin representations of the
ordinary superPoincar\'e algebra. This provides us with a new framework to discuss such representations.
There has been a struggle since the advent of String Theory to describe the zero-tension limit of such
a theory, which should be a theory with massless particles of all possible spins
\cite{fierz,fierzpauli,singh,fronsdal,fronsdal1,Bengtsson:1983pd,Antoniadis:2009rd}. It is hence interesting
to study methods to generate infinite-component massless supermultiplets. In this paper we have only
taken a first step to include half-integer spins in a recently proposed scheme \cite{Savvidy:2010vb}.
In future work we will extend this to higher supersymmetries, higher-dimensional algebras and possibly
further extension along the lines of this paper.

\section*{\it Acknowledgement}
One of us G.S would like to thank CERN Theory Division for hospitality. L.B. wants to thank
Edward Witten for an invitation to the Institute of Advanced Study, where part of the work
was done. This work was supported in part by the European Commission under the ERC Advanced
Grant 226371 and the contract PITN-GA-2009-237920. I. A. was also supported in part by the CNRS grant GRC APIC PICS 3747.

\section*{\it Appendix}

The irreducible representations of the extended algebra can be found by the well-known  method
of induced representations \cite{wigner,Brink:2002zx}. This method consists of finding a representation
of the Wigner's  little group L  and boosting it up to a representation
of the full group. The subgroup L is a group of transformations which leave
a fixed momentum, in our case the time-like momentum $k^{\mu}=k(1,0,0,1)$, invariant.
Under the Lorentz rotations the action of the element
$
U_{\theta}= \exp{({i\over 2} \omega_{\mu\nu} M^{\mu\nu})}
$
creates an infinitesimal transformation $k^{\mu} \rightarrow
\omega^{\mu}_{~\nu} ~k^{\nu} +k^{\mu}$, and
$k^{\mu}=k(1,0,0,1)$ is left invariant provided the parameters obey the relations
$
\omega_{30} =0,~~~\omega_{10} + \omega_{13}=0,~~~\omega_{20} + \omega_{23}=0.
$
Therefore the little subalgebra L contains at least the following generators:
\beqa
h= M_{12},~~~\pi^{'}/P^0= M_{10} + M_{13} ,~~~\pi^{''}/P^0 = M_{20} + M_{23}.\nn
\eeqa
The $M_{12}$ represents the helicity operator $h$:
$$
h = {\vec{P} \vec{S} \over P^0}= {\vec{P} \vec{J} \over P^0}=
{P_i \epsilon_{ijk} M_{jk} \over P^0} = M_{12},
$$
where $( \vec{J} = \vec{R} \times \vec{P} + \vec{S})$.
The super-Poincar\'e little algebra is:
\beqa
&&[h,\pi^{'}]=+
i\pi^{''},~~~~~~~~~[h,\pi^{''}]= -i\pi^{'},~~~~~~~~~~~~~~~[\pi^{'},\pi^{''}]=0,\nn\\
&&[h, ~ \bar{Q}_{\dot{1}} ] = -{1\over 2} \bar{Q}_{\dot{1}} ,~~~~~~[h, ~ Q_1 ]
=  +{1\over 2} Q_1,~~~~~~~~~~~\{Q_1, \bar{Q}_{\dot{1}}\}=4k,\nn \\
&&[h, ~ \bar{Q}_{\dot{2}} ] =  +{1\over 2} \bar{Q}_{\dot{2}} ,~~~~~~[h, ~ Q_2 ]
=  -{1\over 2} Q_2,~~~~~~~~~~~\{Q_2, \bar{Q}_{\dot{2}}\}=0, \\
&&[\pi^{'}, Q_1] =iQ_2,~~ [\pi^{''}, Q_1] = -Q_2,~~[\pi^{'}, \bar{Q}_{\dot{1}}] =i\bar{Q}_{\dot{2}},~~
 [\pi^{''}, \bar{Q}_{\dot{1}}] =  \bar{Q}_{\dot{2}},~~~~\nn\\
 &&[\pi^{'}, Q_2] =0,~~~~~ [\pi^{''}, Q_2] = 0,~~~~[\pi^{'}, \bar{Q}_{\dot{2}}] =0,~~~~~
 [\pi^{''}, \bar{Q}_{\dot{2}}] = 0\nn
\eeqa
and the rest of the anticommutators between the supercharges is
$\{Q_\alpha, Q_\beta,\}=\{\bar{Q}_{\dot{\alpha}}, \bar{Q}_{\dot{\beta}}\}=0$.
The first level commutation relations in (\ref{levelcommutatorsss}) are
$$
 [M^{\mu \nu}, ~ L_{a}^{\lambda }] = i(
\eta^{\lambda \nu } L_{a}^{\mu  }
-\eta^{\lambda\mu} L_{a}^{\nu }),~~~~~~~~~~~[Q_\alpha,L^{\lambda}_{a}] = 0,~~[\bar{Q}_{\dot{\alpha}},L^{\lambda}_{a}] =0
$$
and, when written in components, the little subalgebra L takes the form:
\beqa
\begin{array}{lll}
~[h, ~ L^{0}_{a} ] = [h, ~ L^{3}_{a} ]=0 ~~~~&~[\pi^{'}, ~ L^{0}_{a} ] =-i L_{a}^{1 } ~~~&[\pi^{''}, ~ L^{0}_{a} ] =-i L_{a}^{2 } \\
~[h, ~ L^{0}_{a} ] = [h, ~ L^{3}_{a} ]=0  &~[\pi^{'} , ~ L^{3}_{a} ]=-i L_{a}^{1 } ~~~&[\pi^{''} , ~ L^{3}_{a} ]=-i L_{a}^{2 }    \\
~[h, ~ L_{a}^{1 }] = +i L_{a}^{2 } &~[\pi^{'}, ~ L_{a}^{1 }] = -i (L^{0}_{a}-L_{a}^{3 }) ~~~&[\pi^{''}, ~ L_{a}^{1 }] = 0 \\
~[h, ~ L_{a}^{2 }] = -i L_{a}^{1 } &~ [\pi^{'}, ~ L_{a}^{2 }] = 0 ~~~&[\pi^{''}, ~ L_{a}^{2 }] = -i (L^{0}_{a}-L_{a}^{3 }) .\\
\end{array}       \nn
\eeqa
The second level commutation relations are
$$
 [M^{\mu \nu}, ~ L_{a}^{\lambda_1 \lambda_2}] = i(
\eta^{\lambda_1 \nu } L_{a}^{\mu \lambda_2 }
-\eta^{\lambda_1\mu} L_{a}^{\nu \lambda_2}+
\eta^{\lambda_2 \nu } L_{a}^{\mu \lambda_1 }
-\eta^{\lambda_2\mu} L_{a}^{\nu \lambda_1}),~~~
~[Q_\alpha, L_{a}^{\lambda_1 \lambda_2}] = [\bar{Q}_{\dot{\alpha}},L_{a}^{\lambda_1 \lambda_2}] =0
$$
and in components they have the form:
\beqa\label{secondlevelcomm1}
&&[h, ~ L^{00}_{a} ] = [h, ~ L^{03}_{a} ]=[h, ~ L^{33}_{a} ]=0 \nn \\
&&[h, ~ L_{a}^{01 }] = +i L_{a}^{02 }\nn\\
&&[h, ~ L_{a}^{02 }] = -i L_{a}^{01 }\nn \\
&&[h, ~ L_{a}^{13 }] = +i L_{a}^{23 }\nn\\
&&[h, ~ L_{a}^{23 }] = -i L_{a}^{13 }\nn \\
&&[h, ~ L_{a}^{11 }] = +2 i L_{a}^{12 }\nn\\
&&[h, ~ L_{a}^{22 }] = -2 i L_{a}^{12 }\nn \\
&&[h, ~ L_{a}^{12 }] = +i (L_{a}^{22 }-L_{a}^{11 }),
\eeqa
and with translation operators $\pi^{'}$ and  $\pi^{''}$:
\beqa\label{secondlevelcomm2}
\begin{array}{ll}
~[\pi^{'}, ~ L^{00}_{a} ] =-2i L_{a}^{01 },~~&[\pi^{''}, ~ L^{00}_{a} ] =-2i L_{a}^{02 } \\
~[\pi^{'}, ~ L_{a}^{01 }] = -i L^{00}_{a}+iL_{a}^{03 }-iL^{11}_{a},~~&[\pi^{''}, ~ L_{a}^{01 }] =-i L^{12}_{a} \\
~[\pi^{'}, ~ L_{a}^{02 }] = -i L^{12}_{a} ,~~&[\pi^{''}, ~ L_{a}^{02 }] = -i L^{00}_{a}+iL_{a}^{03 }-iL^{22}_{a} \\
~[\pi^{'} , ~ L^{03}_{a} ]=-i L_{a}^{01 } -i L_{a}^{13 },~~&[\pi^{''} , ~ L^{03}_{a} ]=-i L_{a}^{02 } -i L_{a}^{23 }  \\
~[\pi^{'}, ~ L_{a}^{11 }] = -2iL^{01}_{a} +2i L^{13}_{a},~~&[\pi^{''}, ~ L_{a}^{11 }] = 2i L^{23}_{a}-2iL^{02}_{a} \\
~[\pi^{'}, ~ L_{a}^{12 }] = i L^{23}_{a}-iL^{02}_{a},~~&[\pi^{''}, ~ L_{a}^{12 }] = i L^{13}_{a}-iL^{01}_{a} \\
~[\pi^{'}, ~ L_{a}^{13 }] = -i L^{11}_{a}+iL_{a}^{33 }-iL^{03}_{a},~~&[\pi^{''}, ~ L_{a}^{13 }] = -i L^{12}_{a} \\
~[\pi^{'}, ~ L_{a}^{22 }] =  0,~~&[\pi^{''}, ~ L_{a}^{22 }] =  0 \\
~[\pi^{'}, ~ L_{a}^{23 }] = -i L^{12}_{a},~~&[\pi^{''}, ~ L_{a}^{23 }] = -i L^{22}_{a}+iL_{a}^{33 }-iL^{03}_{a}   \\
~[\pi^{'}, ~ L_{a}^{33 }] =-2i L^{13}_{a},~~&[\pi^{''}, ~ L_{a}^{33 }] =-2 i L^{23}_{a}.
\end{array}
\eeqa
The current subalgebra in (\ref{alakacss} ) between generators has the following form:
\beqa\label{intergalcomm}
&&[L_{a}^{0}, L_{b}^{0}]=if_{abc}
L_{c}^{00},~[L_{a}^{0}, L_{b}^{1}]=if_{abc}
L_{c}^{01},~
 [L_{a}^{0}, L_{b}^{2}]=if_{abc}
L_{c}^{02},~
~[L_{a}^{0}, L_{b}^{3}]=if_{abc}
L_{c}^{03}\nn \\
&&[L_{a}^{1}, L_{b}^{1}]=if_{abc},~
L_{c}^{11},~ [L_{a}^{1}, L_{b}^{2}]=if_{abc}
L_{c}^{12},~ [L_{a}^{1}, L_{b}^{3}]=if_{abc} L_{c}^{13}\nn \\
&&[L_{a}^{2}, L_{b}^{2}]=if_{abc}
L_{c}^{22},~[L_{a}^{2}, L_{b}^{3}]=if_{abc}
L_{c}^{23},~\nn\\
&&[L_{a}^{3}, L_{b}^{3}]=if_{abc}
L_{c}^{33}~.
\eeqa
and so on to the higher levels.

\vfill
\end{document}